\documentclass[superscriptaddress,secnumarabic,amssymb,amsmath,nobibnotes,aps,prd,showkeys,showpacs,twocolumn,nofootinbib]{revtex4-1}%

\usepackage{hyperref}
\usepackage{amsmath}
\usepackage{amssymb}
\usepackage{nicefrac}
\usepackage{graphicx}
\usepackage{color}
\usepackage{subfig}
\usepackage{bm}
\usepackage{gensymb}
\usepackage{bigints}
\usepackage[usenames,dvipsnames,svgnames,table]{xcolor}
\usepackage{pifont}% http://ctan.org/pkg/pifont
\usepackage{aas_macros}
\usepackage{orcidlink}
\usepackage{physics}
\graphicspath{ {./images/} }
\usepackage[normalem]{ulem}
\usepackage{cancel}

\renewcommand{\d}{\mathrm{d}}
\newcommand{\sgn}{\mathrm{sgn}}
\newcommand{\e}{\operatorname{e}}

\usepackage[labelfont=bf]{caption}
\hypersetup{
    colorlinks=true,
    linkcolor=Blue,
    filecolor=Blue,
    urlcolor=MidnightBlue,
    citecolor=Blue,
   pdftitle={Cooling age}
}

\begin{document}

% Use the \preprint command to place your local institutional report
% number in the upper righthand corner of the title page in preprint mode.
% Multiple \preprint commands are allowed.
% Use the 'preprintnumbers' class option to override journal defaults
% to display numbers if necessary
%\preprint{}

%Title of paper
\title{Temperature effects on white dwarfs in modified gravity}
% repeat the \author .. \affiliation  etc. as needed
% \email, \thanks, \homepage, \altaffiliation all apply to the current
% author. Explanatory text should go in the []'s, actual e-mail
% address or url should go in the {}'s for \email and \homepage.
% Please use the appropriate macro for each each type of information

% \affiliation command applies to all authors since the last
% \affiliation command. The \affiliation command should follow the
% other information
% \affiliation can be followed by \email, \homepage, \thanks as well.

\author{Sof\'ia Vidal\orcidlink{0000-0003-0143-0427}}
\thanks{Corresponding author}
\email{sofia.vidal@ut.ee}
\affiliation{Laboratory of Theoretical Physics, Institute of Physics, University of Tartu,
W. Ostwaldi 1, 50411 Tartu, Estonia
}

\author{Aneta Wojnar\orcidlink{0000-0002-1545-1483}}
\email[E-mail: ]{aneta.wojnar@uwr.edu.pl}
\affiliation{Institute of Theoretical Physics, University of Wroc\l aw, pl. Maxa Borna 9, 50-206 Wroc\l aw, Poland}

\author{Laur J\"arv\orcidlink{0000-0001-8879-3890}}
\email[E-mail: ]{laur.jarv@ut.ee}
\affiliation{Laboratory of Theoretical Physics, Institute of Physics, University of Tartu,
W. Ostwaldi 1, 50411 Tartu, Estonia
}

\begin{abstract}
In this article we analyze the effects of a finite temperature equation of state on the equilibrium structure of white dwarfs in massive Brans-Dicke theory as well as the symmetron and dilaton screening mechanisms. We compute and present the numerically obtained mass-radius relation, effective gravitational constant as well as radial profiles of the scalar field, pressure and metric within the star. We show that assuming a non-zero temperature effectively results in a larger radius while leaving the total mass of the star essentially unchanged, and discuss the interplay between the effective gravitational constant, central density, and radius of the star.
\end{abstract}

\maketitle

\section{Introduction}

White dwarfs are compact stellar remnants formed after the evolution of stars with initial masses below roughly ten solar masses \cite{1986bhwd.book.....S,2018MNRAS.480.1547L}. Their interiors are typically composed of carbon and oxygen, although low-mass white dwarfs may possess helium cores, all surrounded by thin outer envelopes of helium and hydrogen. Since thermonuclear reactions have ceased, the evolution of these objects is governed mainly by cooling processes, making white dwarfs important astrophysical laboratories for studying stellar evolution and estimating the ages of galactic populations.

The equilibrium structure of white dwarfs is sustained by electron degeneracy pressure, which counteracts gravitational collapse through the Pauli exclusion principle. This balance leads to the existence of the Anderson–Stoner–Chandrasekhar mass limit \cite{1929ZPhy...56..851A,1930LEDPM...9..944S,1931ApJ....74...81C}, close to $1.4$ solar masses for non-rotating and weakly magnetized configurations. White dwarfs approaching this limit are associated with type Ia supernova explosions, widely used as standard candles in cosmology \cite{choudhuri2010astrophysics,1987ApJ...323..140L,1997Sci...276.1378N}.

Recent observations of unusually over-luminous and under-luminous type Ia supernovae \cite{2006Natur.443..308H,2012ApJ...756..191K,1992AJ....104.1543F,2006A&A...460..793S} suggest that the standard Chandrasekhar picture may not fully describe all compact stellar configurations. These findings have motivated detailed investigations of the nature of SN Ia progenitors \cite{2025A&ARv..33....1R}, as well as a growing interest in alternative scenarios, including modified gravity effects, which could alter the internal structure and maximum mass of white dwarfs \cite{2015JCAP...05..045D,2018JCAP...09..007K,2022PhRvD.105b4028S}. 

Most studies of white dwarfs in modified theories of gravity available in the literature rely on zero-temperature or cold equations of state (see, e.g.\ {\cite{olmo2020stellar,wojnar2021white, 2023PhRvD.107d4072K, PhysRevD.111.084075, Bachs-Esteban:2024vym, universe11050158}}, and references therein), neglecting thermal effects. However, {finite-temperature effects in white dwarfs have been studied extensively within the framework of general relativity} and may significantly affect the stellar structure \cite{2016IJMPS..4160129B}. In principle, these effects can introduce a degeneracy between thermal contributions and deviations caused by modified gravity, similarly to the known degeneracy between the choice of the equation of state and modifications of Einstein’s theory. {Observational constraints on modified gravity parameters using temperature-dependent white dwarf mass-radius measurements have been studied by \cite{Danarianto2023}.}

In this work, we investigate several scalar–tensor gravity models at different temperatures in order to determine whether such a degeneracy could indeed occur and what form it takes.
{Our analysis is intended as a theoretical comparison of equilibrium solutions. The aim of the present work is not to constrain modified gravity theories using white-dwarf observations. The resulting finite-temperature mass-radius relations, scalar-field profiles, metric functions, and effective gravitational constants constitute the main new results presented in this work.} Scalar-tensor theories (STT) pose a prototypical extension of general relativity  \cite{Brans:1961sx,Damour:1992we,CANTATA:2021asi}. Considering a gravitationally minimally coupled scalar field and including quantum corrections, they are derived naturally from compact higher dimensions. A simple approximation to STTs with more complexity is massive Brans-Dicke theory \cite{Brans:1961sx,Damour:1992we}. Astrophysical studies are abundant in this gravity theory family. This includes works on neutron stars (a review is given in \cite{olmo2020stellar,2024RvMP...96a5004D}) as well as white dwarfs \cite{2018JCAP...05..028S,2020PhRvD.101b3017P, 2012MNRAS.423.3328F,2016PhRvL.116o1103J,2022PhRvD.106l4010A,2019PhRvD.100b4025C,2019CQGra..36i5017L,1999MNRAS.305..905B}. Moreover, research on the STT class has been especially thriving as they introduce many phenomenologically successful cosmological models \cite{CANTATA:2021asi}. However, such theories can influence structure formation \cite{PhysRevD.97.081302, PhysRevD.109.023530, WETTERICH2007201, PhysRevD.94.103518} and contradict local gravity tests \cite{will2006theory} as the scalar field introduces an additional, potentially long-range force. This is where the incorporation of screening mechanisms comes in handy. In the symmetron and dilaton case studied here, the field becomes environment dependent, thus reconciling the theory with observations.

We first outline the Chandrasekhar equation of state at finite temperature that describes a degenerate relativistic Fermi gas and its limit for zero temperature in Sec.\ \ref{sec:EOS}. In Sec.\ \ref{sec:hydrostaticequil} we introduce the relativistic field equations for a general, conformal STT in Einstein frame together with the hydrostatic equilibrium equation. Hereafter we present all considered models: massive Brans-Dicke theory, symmetron and dilaton screening. We proceed with a brief description of our numerical setup in Sec.\ \ref{sec:numerics} and a discussion of our numerical results in Sec.\ \ref{sec:results}. Finally, we summarize and conclude in Sec.\ \ref{sec:conclusions}.

\section{Equation of state}\label{sec:EOS}

The matter inside a WD can be considered to be a degenerate relativistic Fermi gas which can be described analytically by the Chandrasekhar equation of state introduced in Ref.\ \cite{1931ApJ....74...81C}.
From kinetic theory, we can write the number density of a specific particle species $n$, energy density $\varepsilon$ and pressure $P$ of a system, respectively, as
\begin{subequations}
    \label{eq:EOS_general}
    \begin{align}
        n &= \frac{g}{h^3}\int \, \d^3p\, f(x, p, t) \;, \label{eq:EOS_numdens_general} \\
        \varepsilon &= \frac{g}{h^3}\int \, \d^3p\, E\, f(x, p, t)\;, \label{eq:EOS_energy_general} \\
        P &= \frac{g}{3h^3}\int \, \d^3p\, pv\, f(x, p, t)\;,  \label{eq:EOS_pressure_general}
    \end{align}
\end{subequations}
{where $f({x},{p},t)$ is the distribution function in phase
space, ${x}$ the spatial coordinate, ${p}$ the momentum,
$t$ the time, $g$ the degeneracy number, and $h$ Planck's constant.
The total relativistic single-particle energy is
$E_{\rm tot}(p)=\sqrt{p^2c^2+m^2c^4}$, and we assume an isotropic
momentum distribution with velocity $v=pc^2/E_{\rm tot}$.
For the Fermi-Dirac distribution, it is convenient to introduce the
kinetic energy and kinetic chemical potential,
\begin{equation}
E(p) = E_{\rm tot}(p)-m_ec^2, \qquad
\mu = \mu_{\rm tot}-m_ec^2.
\end{equation}
The distribution function for electrons can then be written equivalently as
\begin{equation}
f(p)
=
\frac{1}{
\exp\left[
\frac{E_{\rm tot}(p)-\mu_{\rm tot}}{k_B T}
\right]+1
}
=
\frac{1}{
\exp\left[
\frac{E(p)-\mu}{k_B T}
\right]+1
},
\end{equation}
with Boltzmann constant $k_B$ and temperature $T$.}
Hence, we can rewrite the quantities given above \eqref{eq:EOS_general} as follows
\begin{subequations}
    \label{eq:EOS_FermiDirac}
    {\begin{align}
        n_e &= \frac{2}{h^3}\int_0^\infty \, \frac{4\pi p^2\,\d p}{\exp\left[\frac{E(p)-\mu}{k_BT}\right]+1} \;, \label{eq:EOS_numdens_FD} \\
        \varepsilon &= \frac{2}{h^3}\int_0^\infty \, \frac{4\pi p^2\, E(p)\,\d p}{\exp\left[\frac{E(p)-\mu}{k_BT}\right]+1} \;, \label{eq:EOS_energy_FD} \\
        \label{eq:EOS_pressure_FD} 
        P &= \frac{2}{3h^3}\int_0^\infty \, \frac{p^2c^2}{E(p)} \frac{4\pi p^2\d p}{\exp\left[\frac{E(p)-\mu}{k_BT}\right]+1} \;, 
    \end{align}}%
\end{subequations}%
using $g_e = 2$ for electrons. We will further assume a carbon-oxygen composition for the stellar core in this work.

\subsection{Finite temperatures}

While the energy density is dominated by the rest mass energy density, and can be thus safely approximated to consist only of this term, the largest pressure contribution results from the degenerate electrons. In the present analysis, we neglect the ionic and radiation contributions to the pressure. The electron pressure gives the dominant contribution in most of the WD interior. However, this approximation becomes less accurate in the outer layers, where the density is lower, especially for the highest considered temperature, $T=10^8$ K. In this regime, the ionic and radiation pressures become non-negligible. Therefore, the results obtained for such high temperatures should be rather treated as a qualitative illustration of the finite-temperature effects in the electron equation of state, and not as a complete thermodynamic description of a hot WD.

Equation \eqref{eq:EOS_numdens_FD} can be reformulated as
\begin{equation}
    n_e (\beta, T) = \frac{\sqrt{2}m_e^3c^3}{\pi^2\hbar^3}\beta^{\nicefrac{3}{2}}\left[F_{\nicefrac{1}{2}}(\eta, \beta) + \beta F_{\nicefrac{3}{2}}(\eta, \beta)\right] \;,
    \label{eq:EOS_numdens_finitetemp}
\end{equation}
with $\hbar = h/2\pi$ and using relativistic Fermi-Dirac integrals
\begin{equation}
    F_{k}(\eta, \beta) = \int_0^\infty \, \frac{t^k\sqrt{1+\frac{1}{2}\beta t}\,\d t}{1+\exp(t-\eta)} \;,
\end{equation}
where we have defined the degeneracy parameters ${\eta = \mu/(k_BT)}$, ${t=E(p)/(k_BT)}$ and ${\beta = k_BT/(m_ec^2)}$.
Hence, the pressure becomes
\begin{equation}
    P (\beta, T) =  \frac{{2}^{\nicefrac{3}{2}}m_e^4c^5}{3\pi^2\hbar^3}\beta^{5/2}\left[F_{\nicefrac{3}{2}}(\eta, \beta) + \frac{\beta}{2} F_{\nicefrac{5}{2}}(\eta, \beta)\right]\:.
    \label{eq:EOS_pressure_finitetemp}
\end{equation}
{To construct the equation of state $P(\rho,T)$, we relate the electron
number density to the mass density through charge neutrality,
\begin{equation}
n_e = \frac{\rho}{\mu_e m_u},
\end{equation}
where $m_u$ is the atomic mass unit and $\mu_e=A/Z$ is the mean
molecular weight per electron. For the carbon-oxygen composition
considered here, $\mu_e=2$, and hence $n_e=\rho/(2m_u)$. For each pair
$(\rho,T)$, Eq.~(4) is solved for the degeneracy parameter $\eta$,
which is subsequently inserted into Eq.~(6) to obtain $P(\rho,T)$.}

\begin{figure}
    \centering
    \includegraphics[width=.95\linewidth]{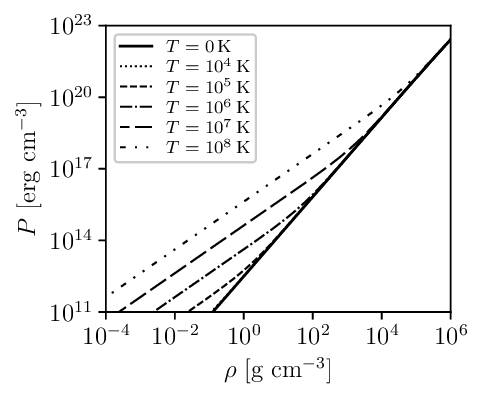}
    \caption{Temperature dependent Chandrasekhar equation of state in cgs units for $T\in [0, 10^8]$ K.}
    \label{fig:EOS}
\end{figure}

{The effect of temperatures on the equation of state is shown in Fig.\ \ref{fig:EOS}. The density range is restricted to $\rho \le 10^6\,\mathrm{g\,cm^{-3}}$ since finite temperature effects become negligible at higher densities. Above this threshold, the equation of state is practically indistinguishable from the zero-temperature limit.}

{Throughout this work, $T$ denotes the local matter temperature entering
the equation of state, which we approximate as spatially constant within
each stellar configuration. Strict global thermal equilibrium would
instead require the Tolman relation in Jordan frame
\begin{equation}
T(r)\sqrt{-g_{tt}(r)} = T_{\infty}.
\end{equation}
Since white dwarfs have low compactness,
${GM/(Rc^2) \ll 1}$, the corresponding variation of the local temperature
is small and is neglected here.}

\subsection{Zero temperature limit}

For a fully degenerate gas, i.e. ${T \to 0K}$, the distribution function reduces to

\begin{equation}
    f(E) = \begin{cases}
    1,&  E\leq E_F\\
    0,&  E> E_F\;,
    \end{cases}
    \label{eq:EOS_distributionfunction_nulltemp}
\end{equation}
where we have defined the Fermi energy ${E_F:= \mu(T\to 0)}$ that we can be related to the Fermi momentum $p_F$ set by {${E_F = \sqrt{p_F^2c^2+m_e^2c^4}}-m_ec^2$}.

Consequently, the electron number density is
\begin{equation}
    n_e = \frac{2}{h^3}\int_0^{p_F} \, 4\pi p^2\,\d p = \frac{8\pi p_F^3}{3h^3}
\end{equation}
and the pressure
\begin{equation}
    P(x) = \frac{\pi m_e^4 c^5}{3h^3} \left[(2x^3-3x)\sqrt{1 + x^2} + 3\sinh^{-1}{x} \right]\;, \label{eq:EOS_pressure_zerotemp}
\end{equation}
where $x=p_F/m_ec$ represents the dimensionless Fermi momentum.

{As a consistency check, we verified that the finite-temperature equation
of state continuously approaches the standard Chandrasekhar
zero-temperature result as $T \to 0$.}

\section{Hydrostatic Equilibrium Equations} \label{sec:hydrostaticequil}

In the following, we introduce the hydrostatic equilibrium equations for scalar-tensor theories in Einstein frame. This frame is obtained from the physical Jordan frame by a conformal transformation of the metric tensor, written as $g_{\mu\nu} = A^2(\tilde{\varphi})\tilde{g}_{\mu\nu}$ and scalar field redefinition. We refer the reader to Refs.\ \cite{Damour:1992we,Doneva:2013qva,2024RvMP...96a5004D,Jarv:2014hma} for a more detailed analysis of these two frames. For clarity, quantities with a tilde, e.g.\ $\tilde r$, will represent Einstein frame quantities, whereas quantities without, e.g.\ $r$, are the physical ones in Jordan frame.

Scalar-tensor theories (STTs) can generally be described by an Einstein frame action of the form
\begin{align}\label{eq:action}
    S =& \frac{{c^4}}{16\pi G}\int \frac{\d^4x}{{c}} \sqrt{-\tilde{g}} \left[\tilde{R} - 2\tilde{g}^{\mu\nu}\tilde\partial_{\mu}\tilde{\varphi}\tilde\partial_{\nu}\tilde{\varphi}  - V(\tilde{\varphi})\right] \nonumber \\
    &+ S_\mathrm{m}\left(A^2(\tilde{\varphi})\tilde{g}_{\mu\nu}, \chi\right)\;,
\end{align}
where $c$ is the speed of light and $G$ the bare gravitational constant. The specific STT is then defined by the choice of conformal factor $A(\tilde{\varphi})$ and self-interacting potential $V(\tilde{\varphi})$. Here, our Jordan frame field $\phi$ and Einstein frame field $\tilde\varphi$ are related via $\phi = A^{-2}(\tilde{\varphi})$.

The field equations follow by varying the action with respect to the metric and scalar field
\begin{align}
    \tilde{R}_{\mu\nu} - \frac{1}{2}\tilde{g}_{\mu\nu}\tilde{R} =&\; \frac{8\pi G}{c^4} \mathcal{\tilde{T}}_{\mu\nu} + 2\tilde\nabla_{\mu}\tilde{\varphi}\tilde\nabla_{\nu}\tilde{\varphi} \nonumber \\
    &-\tilde g_{\mu\nu}\tilde g^{\alpha\beta}\tilde\nabla_{\alpha}\tilde{\varphi}\tilde\nabla_{\beta}\tilde{\varphi}-\frac{1}{2}V(\tilde{\varphi})\tilde{g}_{\mu\nu}\;, \label{eq:FE_metric} \\
    \tilde\nabla_{\mu}\tilde\nabla^{\mu}\tilde{\varphi} =& -\frac{4\pi G}{{c^4}}\alpha(\tilde{\varphi})\mathcal{\tilde{T}} + \frac{1}{4}\frac{\d V(\tilde{\varphi})}{\d\tilde{\varphi}} \,.
    \label{eq:FE_scalarfield}
\end{align}
We assume the energy-momentum tensor of a perfect fluid $\mathcal{\tilde T_{\mu\nu}}$ that relates to the Jordan frame tensor through $\mathcal{\tilde T_{\mu\nu}} = A^2(\tilde{\varphi})\mathcal{T_{\mu\nu}}$. Hence, the relations between physical and Einstein frame thermodynamic variables, pressure and density, are $\tilde{P} = A^4(\tilde{\varphi})P$ and $\tilde{\rho} = A^4(\tilde{\varphi})\rho$, respectively. We have further introduced the logarithmic derivative of the conformal factor with respect to the scalar field ${\alpha(\tilde{\varphi}) = \frac{\d\ln{A(\tilde{\varphi})}}{\d\tilde{\varphi}}}$. This function reflects the coupling strength between the field and matter.

Assuming a spherically symmetric and static configuration of the interior structure of the WD, we can set the metric to
\begin{equation}\label{eq:metricBD}
    \d \tilde{s}^2 = -\e^{2\tilde{\phi}(r)}c^2\d t^2 + \e^{2\tilde{\Lambda}(r)}\d \tilde{r}^2 + \tilde{r}^2\d\Omega^2\;,
\end{equation}
where ${\Omega^2 = \d\theta^2 + \sin^2{\theta}\d\vartheta^2}$ is the angular element, reducing the field equations to
{\small{\begin{subequations}
    \label{eq:FE_sphsymm}
    \begin{align}
        \frac{2\tilde{\Lambda}'}{\tilde{r}} &= \frac{8\pi G}{c^2} A^4(\tilde{\varphi})\rho\e^{2\tilde{\Lambda}} + \frac{1-\e^{2\tilde{\Lambda}}}{\tilde{r}^2} +\tilde{\varphi}'^2 +\frac{V(\tilde{\varphi})}{2}\e^{2\tilde{\Lambda}}\;, \label{eq:FE_sphsymm_lambda} \\
        \frac{2\tilde{\phi}'}{\tilde{r}} &= \frac{8\pi G}{c^4} A^4(\tilde{\varphi})P\e^{2\tilde{\Lambda}} - \frac{1-\e^{2\tilde{\Lambda}}}{\tilde{r}^2} +\tilde{\varphi}'^2-\frac{V(\tilde{\varphi})}{2}\e^{2\tilde{\Lambda}}\;, \label{eq:FE_sphsymm_phi} \\
        \tilde{\varphi}'' &+ \left(\tilde{\phi}'-\tilde{\Lambda}'+\frac{2}{\tilde{r}}\right)\tilde{\varphi}' \nonumber\\
        &\quad\quad\quad= \left[\frac{4\pi G}{{c^4}}\alpha A^4(\tilde{\varphi})(\rho c^2-3P) + \frac{1}{4}\frac{\d V}{\d\tilde{\varphi}}\right]\e^{2\tilde{\Lambda}}\;, \label{eq:FE_sphsymm_varphi}
    \end{align}
\end{subequations}}}
where derivatives with respect to the radius $\tilde{r}$ are indicated by primes. {Note that thermodynamical variables, pressure $P$ and density $\rho$, are already given in Jordan frame in the equations above and will be throughout this manuscript.}

The system of differential equations is then closed with the equation for hydrostatic equilibrium
\begin{equation}
    P' = -(\rho c^2 + P)\left(\tilde{\phi}'+\alpha\tilde{\varphi}' \right)\;.\label{eq:hydrostaticequilibrium}
\end{equation}
as well as the equation of state introduced above associating the physical pressure $P$ to the density $\rho$.
The central pressure $P_c$ is a free input parameter, such that we have the boundary conditions
\begin{equation*}
    P(\tilde{r}=0) = P_c \quad\mathrm{and}\quad \tilde{\Lambda}(\tilde{r}=0) = 0\;,
\end{equation*}
at the stellar center. {For massive Brans-Dicke theory} the spatial asymptotics {are}
\begin{equation*}
    \lim_{\tilde{r}\to\infty} \tilde{\Lambda} (\tilde{r}) = 0  \quad\mathrm{and}\quad \lim_{\tilde{r}\to\infty} \tilde{\phi}(\tilde{r}) = 0\;,
\end{equation*}
for the metric functions {while in the screening cases we obtain a Schwarzschild-de Sitter spacetime}. The asymptotic of the scalar field depends on the theory taken into consideration. Outside of the star, the trace of the energy momentum tensor vanishes ${\mathcal{\tilde T} = 0}$, reducing the scalar field equation \eqref{eq:FE_scalarfield} to ${\tilde\Box\tilde{\varphi} \propto \frac{\d V(\tilde{\varphi})}{\d\tilde{\varphi}}}$. Thus, the asymptotic value of the scalar field is set by the potential minimum.

The radius of the star $\tilde{r}_s$ in Einstein frame is the smallest radius that satisfies the condition $P(\tilde{r}_s) = 0$, while the physical stellar radius is $R_s = A[\tilde{\varphi}(\tilde{r}_s)]\tilde{r}_s$.

Below, we briefly present the theories studied in this work. A more detailed description of the relevant theories can be found in, e.g.\, Refs. \cite{Brans:1961sx,Damour:1992we} for massive Brans-Dicke theory, Ref. \cite{PhysRevLett.104.231301} for the symmetron and Ref. \cite{PhysRevD.82.063519} for the dilaton screening mechanism.

\subsection{Massive Brans-Dicke Theory}

In the original formulation of Brans and Dicke \cite{Brans:1961sx}, the coupling strength $\alpha(\tilde\varphi)$ is assumed to be field-independent and constant. Therefore, we can write
\begin{equation}
    A(\tilde{\varphi}) = \e^{\alpha_0\tilde{\varphi}}\;, \quad\mathrm{hence}\quad \alpha(\tilde{\varphi})= \alpha_0\;,
    \label{eq:BD_conffactor}
\end{equation}
and we further introduce the simple self-interacting potential
\begin{equation}
    V(\tilde{\varphi}) = \frac{2m^2_{\tilde{\varphi}}c^2}{\hbar^2}\tilde{\varphi}^2\;.
    \label{eq:BD_potential}
\end{equation}
The scalar field mass $m_{\tilde{\varphi}}$ effectively reduces the range of the scalar field to its Compton wavelength ${\lambda_{\tilde\varphi}= 2\pi{\hbar}/m_{\tilde{\varphi}}{c}}$. It is clearly visible that the effect of the field will be suppressed until negligible over a shorter distance the more massive it is. At this length, the Yukawa correction coming from the field is eliminated from the Newtonian gravitational potential \cite{Jarv:2014hma}. The scalar field asymptotic is consequently given by
\begin{equation*}
    \lim_{\tilde{r}\to\infty} \tilde{\varphi}(\tilde{r}) = 0\;.
\end{equation*}
This toy model is widely used as a simple approximation of more elaborate STTs \cite{olmo2020stellar,2024RvMP...96a5004D,1999MNRAS.305..905B,2012MNRAS.423.3328F} as it embodies the key characteristics of such theories. It is sufficient for our purpose of studying how the inclusion of temperature in the EOS may influence the structure of WDs in STTs. It is still important to note that this theory does not reduce to GR for a constant scalar field. GR is only obtained for $\alpha_0 = 0$.

Note that massless Brans-Dicke theory, i.e.\ $m_{\tilde{\varphi}} = 0$, is highly constrained by cosmological and astrophysical observations to ${\alpha_0 \lesssim 10^{-5}}$ \cite{2012MNRAS.423.3328F}. As a massive field limits its interaction range, this restriction is softened in the massive case. The parameter $\alpha_0$ can basically assume any value for scalar field masses ${m_{\tilde{\varphi}}c^2 \gtrsim 2\times 10^{-16}}$ eV \cite{Perivolaropoulos:2009ak}. Here, we assume values in the range ${m_{\tilde{\varphi}}c^2\; \in [1.3\times 10^{-13},\; 2.7 \times 10^{-12}]}$ eV and ${\alpha_0 = 1}$ which are well within the observational constraints.

\subsection{Screening Mechanisms}

The scalar field's strength becomes environment dependent when screening mechanisms are incorporated into STTs. In the symmetron and dilaton case particularly, how matter couples to the scalar field depends on the environment. Looking at equation \eqref{eq:FE_scalarfield}, we can define an effective potential that governs the behavior of the field as follows
\begin{equation}
    \frac{\d V_\mathrm{eff}(\tilde\varphi)}{\d\tilde\varphi} = -\frac{4\pi G}{c^4}\alpha(\tilde\varphi)\tilde{\mathcal{T}} + \frac{1}{4}\frac{\d V(\tilde\varphi)}{\d\tilde\varphi}\;.
    \label{eq:screening_effectivepotential}
\end{equation}
Effectively, this renders the scalar field negligible on astrophysical scales, while it still may play a role on cosmological scales. Within compact objects, screening mechanisms partially break down. Hence, stellar properties are modified when considering such theories.

In the following, we introduce the theory behind symmetron and dilaton screening. Note that our notation in \eqref{eq:action} differs from the usual convention used for screening mechanisms, e.g.\ in Refs.\ \cite{PhysRevLett.104.231301, PhysRevD.82.063519, universe11050158}. Nevertheless, we maintain our convention for consistency within this article and with our previous work \cite{PhysRevD.111.084075}. We refer the reader to the appendix \ref{append:notation} for the correspondence between the two formulations.

\subsubsection{Symmetron Screening}

In symmetron screening, the field is suppressed in regions of high density - such as a compact star -, whereas its value and effect grow in low density regions. While the former case preserves the reflection symmetry ${\tilde\varphi \to -\tilde\varphi}$, it is spontaneously broken in the latter. Thus, the name \textit{symmetron}.

In this mechanism, the matter coupling is set to
\begin{equation}
    A(\tilde{\varphi}) = 1 + \frac{\tilde{\varphi}^2}{\bar{M}_S^2} \;.
    \label{eq:Symm_conffactor}
\end{equation}
The theory parameter $\bar{M}_S = M_S/M_P$ is dimensionless since $M_S$ has mass dimension and $M_P = \sqrt{\hbar c / 8\pi G}$ is the Planck mass. The symmetry-breaking potential is
\begin{equation}
    V(\tilde{\varphi}) = -2\mu^2\tilde\varphi^2 + \lambdabar \tilde\varphi^4
    \label{eq:Symm_potential}
\end{equation}
where $\mu$ has mass dimension and $\lambdabar = {2}\lambda M_P^2$ with a dimensionless parameter $\lambda$.

As $\tilde\varphi/\bar M_S$ is small and the trace of the energy momentum tensor is dominated by density, we can approximate $\tilde{\mathcal{T}} = A^4(\tilde\varphi)\left(3P - \rho{c^2}\right)\approx -\rho{c^2}$. Thus, the effective potential \eqref{eq:screening_effectivepotential} is given by
\begin{equation}
    V_\mathrm{eff} (\tilde\varphi) \approx \frac{1}{2}\left(\frac{8\pi G}{c^4}\frac{\rho{c^2}}{\bar M_S^2}  - \mu^2\right)\tilde\varphi^2 + \frac{1}{4}\lambdabar\tilde\varphi^4\;.
\end{equation}
The quadratic term is negative or positive depending on the local density. In regions of high enough mass density, the density term dominates over $\mu$ and the vacuum expectation value of the scalar field settles around ${\tilde\varphi = 0}$. In contrast, in low density regions $\mu$ dominates over the mass density term, driving the scalar field towards ${\tilde\varphi = \mu/\sqrt{\lambdabar}}$.
Hence, our scalar field asymptotics are
\begin{equation*}
    \lim_{\tilde{r}\to\infty} \tilde{\varphi}(\tilde{r}) = \frac{\mu}{\sqrt{\lambdabar}}\;.
\end{equation*}

\subsubsection{Dilaton Screening}

In the dilaton case, it is the coupling function that drives the screening mechanism. 
First, the minimum of the conformal factor is set to $\tilde\varphi_d$. Introducing the dimensionless, positive coupling parameter $a_2$, we can thus write $A(\tilde\varphi)$ as
\begin{equation}
    A(\tilde{\varphi}) = 1 + a_2\left(\tilde{\varphi} - \tilde{\varphi}_d\right)^2 \;.
    \label{eq:Dilaton_conffactor}
\end{equation}
Looking at the coupling strength {$\alpha(\tilde\varphi)$ in linear approximation as proportional to} $ a_2\left(\tilde{\varphi} - \tilde{\varphi}_d\right)$, we can see that the force mediated by the field vanishes at the minimum $\tilde\varphi_d$ while the field itself can still be present.

The potential on the other hand follows an exponential runaway behavior
\begin{equation}
    V(\tilde{\varphi}) = 2A^4(\tilde{\varphi})V_0 \e^{-\sqrt{2}\left(\tilde{\varphi}-\tilde{\varphi}_d\right)}.
    \label{eq:Dilaton_potential}
\end{equation}
The potential parameter has mass squared dimension and can be written as $V_0 = \bar V_0 M_P^2$ with a dimensionless parameter $\bar V_0$. Note that the potential minimum is still affected by the conformal factor $A(\tilde\varphi)$. Thus, using the same approximations as above, the effective potential is
\begin{equation}
    V_\mathrm{eff}(\tilde\varphi) \approx \frac{8\pi G}{c^4}\ln{A(\tilde\varphi)}\rho{c^2} + V(\tilde\varphi)\;.
\end{equation}
In high density regions, the first term dominates and the field converges towards the minimum of the coupling function, while for low densities, the field settles into the potential minimum. Finally, the scalar field asymptotic is given by
\begin{equation}
    \lim_{\tilde{r}\to\infty} \tilde{\varphi}(\tilde{r}) {= \tilde{\varphi}_d  + \sqrt{8} - \sqrt{8-{a_2}^{-1}}} \approx \tilde{\varphi}_d + \frac{\sqrt{2}}{8a_2} \;,
\end{equation}
{where we used ${\sqrt{1+x} \approx 1+\frac{1}{2}x}$ for ${x=1/8a_2 < 1}$ since ${a_2 \geq 10}$ in the last step. This approximation was tested for against the full expression for the smallest value of $a_2 = 10$. We did not obtain any differences in the results, thus the approximation was used for larger values of $a_2.$}

\section{Numerical Preliminaries} \label{sec:numerics}

Here, we briefly describe the numerical setup. For a more detailed description of our procedure, we refer the reader to our previous work \cite{PhysRevD.111.084075}.

The system of differential equations \eqref{eq:FE_sphsymm}, \eqref{eq:hydrostaticequilibrium} poses a boundary value problem as the boundary conditions are given at the center of the star as well as at the spatial asymptotics. In order to solve this numerically, a shooting method is implemented to transform it into an initial value problem. In our case, the only shooting parameter needed is the scalar field. The temporal metric function ${\tilde\phi(\tilde r)}$ enters the equations only through its radial derivative ${\tilde\phi'(\tilde r)}$. Thus, we can safely eliminate it from the equations and do not need to consider it as an additional parameter. A comprehensive explanation of the shooting method is given in appendix \ref{append:shootingmethod}. {We further implement an adaptive Runge-Kutta method to integrate the system of differential equations. The scalar-field boundary condition at infinity is satisfied with a tolerance of $10^{-10}$. The integration domain is adjusted according to the considered theory parameters and extends sufficiently far into the asymptotic regime to ensure accurate mass extraction.}
%Numerical convergence has been verified by varying the integration tolerance.
Further, in natural units ${G, \hbar, c = 1}$ and we set ${10^7M_\odot = 1}$. We adopt dimensionless radial variables, density, and pressure. Dimensionfull parameters which have units of some power of mass are transformed to dimensionless parameters with
\begin{equation*}
    m_{\mathrm{dim}} = \frac{{c\hbar}}{10^7M_\odot\; G} \approx 2.38 \times 10^{-50} g\;.
\end{equation*}

The numerical condition defining the stellar surface is ${P(\tilde{r}_s) = 10^{-10}P_c}$, i.e. the radius at which the pressure has dropped by ten orders of magnitude from its central value. Note that demanding more orders of magnitude does not significantly change the obtained radii. Varying this value an order of magnitude higher or lower leads to a median relative deviation of approximately $0.39\%$ at $T=10^7 K$ for selected theory parameters across all theories studied. Deviations for smaller temperatures are even lower. At $T=10^8 K$ we obtain a median of $\approx 2.7\%$. Though this deviation is non-negligible, as mentioned above, thie high temperature case is an exploratory limit.
Hence, we can safely assume this to be equivalent to the theoretical condition ${P(\tilde{r}_s) = 0}$. 
{Finally, in the massive Brans-Dicke case, we obtain the stellar mass is via the radial metric asymptotics}

\begin{equation*}
        {M = \frac{r_\infty}{2} \left[1 - \e^{-2\Lambda\left(r_\infty\right)}\right]} \;,
\end{equation*}
{where we use the radius $r_\infty$ at numerical spatial infinity. For both screening mechanisms, we integrate the mass equation according to \cite{universe11050158} to facilitate comparison. All stellar masses obtained are in Jordan frame, i.e.\ the physical ones.}

\section{Results} \label{sec:results}

In this section we present and discuss the obtained mass-radius relations for each considered gravity theory, i.e.\ massive Brans-Dicke theory, symmetron and dilaton screening, at different finite temperatures. These relations describe the equilibrium configurations for WDs in a specific theory, meaning the configuration for which the pressure exactly counterbalances gravity or vise versa. They are acquired through the integration of \eqref{eq:FE_sphsymm}, \eqref{eq:hydrostaticequilibrium} with the respective conformal factor ${A(\tilde\varphi)}$, the potential ${V(\tilde\varphi)}$ and the boundary conditions for each theory presented above in \ref{sec:hydrostaticequil}.
We further review the effect of temperature on the interior radial profile of some physical quantities, such as scalar field and radial metric function $\tilde\Lambda(\tilde r)$, with specific examples for each theory. {We would like to stress that for the results presented below mass $M$, radius $r$, effective gravitational constant $G_\mathrm{eff}$, pressure $P$ and density $\rho$ are Jordan frame quantities. On the other hand, the scalar field $\tilde\varphi$, its derivative $\tilde\varphi'$ and the radial metric function $\tilde\Lambda$ are given in Einstein frame.}

All results are computed for temperatures $T \in [10^4, 10^8]K$. Note that we also computed all the results presented here for $T=0K$. As these show no significant difference from the results for $T=10^4K$ we do not show them separately in the following figures. {The mean
absolute relative difference between these two curves is $0.14\%$ with a maximum relative divergence of $1.14\%$}.
We consider central densities within the range $\rho_c \in [10^5, 10^{12}]\, \mathrm{g}\,\mathrm{cm}^{-3}$.

{The relativistic Fermi-Dirac description itself remains applicable when the electrons become less degenerate. However, at sufficiently high temperatures and low densities, the simplified matter model adopted here, in which the pressure is dominated by the electron contribution, becomes progressively less accurate. In particular, ionic and radiative contributions, as well as Coulomb, ionization, and envelope effects, may no longer be negligible. The $T=10^8$K sequences should therefore be regarded as an exploratory limiting case rather than as a fully realistic model of hot white dwarfs.}

\begin{figure*}
    \centering
    \subfloat[General Relativity for different temperatures. \label{fig:massradius_GR}]{
        \includegraphics[width=0.45\textwidth]{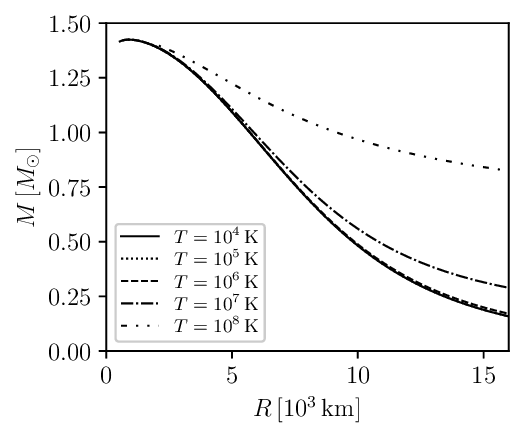}
    }\hfill
    \subfloat[Massive Brans-Dicke theory with $\alpha_0 = 1$ and dimensionless field masses ${m_{\tilde{\varphi}}[1] \in [1\times 10^{4}, 2 \times 10^{5}]}$.  \label{fig:massradius_BD}]{
        \includegraphics[width=0.45\textwidth]{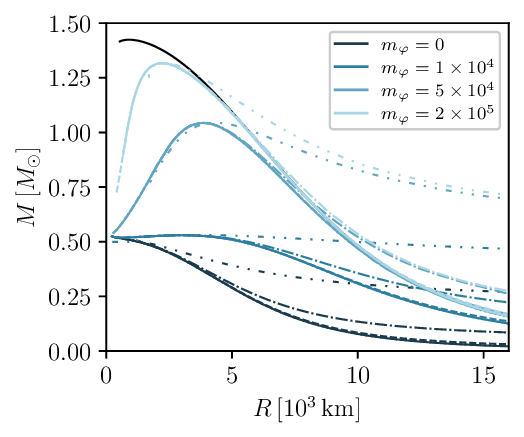}
    }

    \medskip

    \subfloat[Symmetron Screening for ${\mu [1] \in \{1.8, 5.4, 9.0\}\times 10^{3}}$, ${\bar{M}_S=10^{-2}}$, and ${\sqrt{\lambda} = \sqrt{2}\mu[1] /\bar M_S^2}$.  \label{fig:massradius_symmetron}]{
        \includegraphics[width=0.45\textwidth]{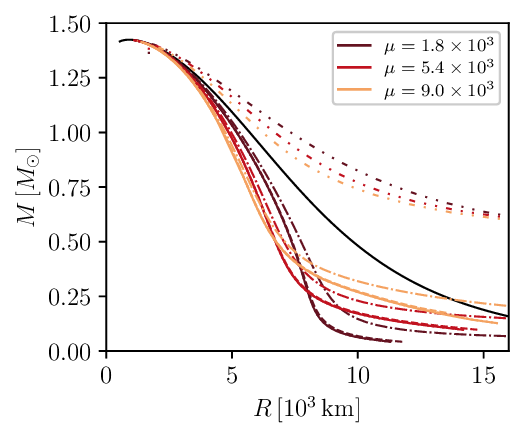}
    }\hfill
    \subfloat[Dilaton screening with $\tilde\varphi_d = 0$, $\log_{10}(a_2) \in \{1, 2, 3, 4\}$ and ${V_0 [1] \in 3.3211 \times [10^{-3}, 1]}$. \label{fig:massradius_dilaton}]{
        \includegraphics[width=0.45\textwidth]{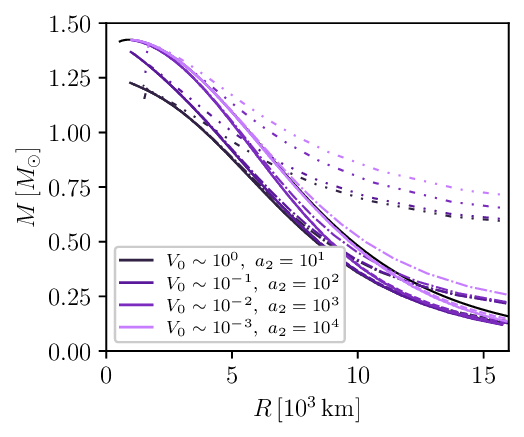}
    }

    \caption{The stellar mass as a function of radius for WDs in GR (top left), massive Brans-Dicke theory (top right), symmetron (bottom left) and dilaton screening (bottom right). The GR curve is plotted for comparison for all modified theories (black solid line). Results are presented {in the physical Jordan frame and} for temperatures ${T \in [10^4, 10^8]K}$. All plots follow the temperature linestyle legend given in Subfig.\ \ref{fig:massradius_GR}. Note that the zero temperature curves essentially do not differ from the curve for ${T=10^4K}$. Thus, they are not explicitly included. This is the case for all results plotted in this paper.}
    \label{fig:massradius}
\end{figure*}

\subsection{Mass-radius relations} \label{subsec:results_massradius}

We obtain the mass-radius relation for each theory by solving the field equations \eqref{eq:FE_sphsymm} together with the hydrostatic equilibrium equation \eqref{eq:hydrostaticequilibrium} and the equation of state \eqref{eq:EOS_numdens_finitetemp}, \eqref{eq:EOS_pressure_finitetemp} with the corresponding conformal factor, self-interacting potential and boundary conditions for the scalar field.
{The presented sequences correspond to equilibrium solutions only. The dynamical stability of configurations beyond the maximum-mass point is not investigated in this work.}
Before going to the modified gravity theories, we note that we also computed the mass-radius relations with a temperature dependent equation of state for GR. {The results are shown in Fig.\ \ref{fig:massradius_GR} and are in good qualitative agreement with the finite-temperature white dwarf sequences presented by \cite{2016IJMPS..4160129B}. In particular, we recover the increase of the stellar radius with increasing temperature, while the maximum mass remains largely unaffected.}

In the massive Brans-Dicke case, we assume the conformal factor and self-interacting potential given in \eqref{eq:BD_conffactor} and \eqref{eq:BD_potential} respectively. For the theory parameters, we assume $\alpha_0 = 1$ and dimensionless masses ${m_{\tilde{\varphi}}[1] \in [1\times 10^{4}, 2 \times 10^{5}]}$ which corresponds to ${m_{\tilde{\varphi}}c^2\;[eV] \in [1.3\times 10^{-13},\; 2.7 \times 10^{-12}]}$ or equivalently ${m_{\tilde{\varphi}}\;[g] \in [2.4\times 10^{-46},\; 4.7 \times 10^{-45}]}$.
The numerically obtained curves for massive Brans-Dicke theory are presented in Fig.\ \ref{fig:massradius_BD}. We observe the same qualitative behavior as for GR. Up to temperatures of $\sim 10^6K$ there is almost no difference in the resulting mass-radius relations. While above those the deviation becomes increasingly significant with higher temperature. Hot stars have larger radii compared to cooler white dwarfs when looking at the same total mass. This effect arises from the increasing thermal pressure at higher temperatures that forces the star to expand.

{For both screening mechanisms, the parameter values adopted in this work coincide with those used in Ref.\ \cite{universe11050158}, allowing a direct comparison with previous studies of white dwarfs in symmetron and dilaton gravity. Our aim is not a comprehensive exploration of the currently allowed parameter space but rather to investigate how finite temperature effects modify equilibrium configurations previously studied in the literature.}
Explicitly, the parameters we adopt for the symmetron are ${\mu\, [1] \in \{1.8, 5.4, 9.0\}\times 10^{3}}$ corresponding to ${\mu\, [M_P] \in [1, 5]\times 10^{-41}}$, ${\bar{M}_S = 10^{-2}}$, and ${\sqrt{\lambda} = \sqrt{2}\mu M_P/M_S^2}$. These values are obtained from the thin-shell factor to ensure that a typical white dwarf is indeed screened. Note that observational bounds on the other hand, demand ${\bar{M}_S \lesssim 10^{-4}}$ and ${\mu\, [M_P] \gtrsim 10^{-56}}$ \cite{PhysRevD.84.103521}. As expected, the same effects as for massive Brans-Dicke theory are recognized for the symmetron, Fig.\ \ref{fig:massradius_symmetron}. At the same mass, a hotter star finds its equilibrium at larger radii as thermal pressure forces expansion.  This also leads to a partial overlap of theories with different sets of parameters. The mass-radius curves become degenerate, making it more complex to distinguish between different theory parameters or a different stellar temperature. Further, we note that the curves for ${T=10^8K}$ seem to converge towards larger radii for all theories towards the same value. This might hint at temperature effects dominating over modified gravity effects in this regime. Still, large radii correspond to low central densities. We are thus close to values that are not realistic for white dwarfs anymore and where the finite temperature EoS looses its validity as the gas becomes less and less degenerate. So this observation need to be taken in with care. We also observe a lower Chandrasekhar mass and sharp drop-off thereafter for the symmetron and dilaton at ${T=10^8K}$. At such high temperatures, we are approaching the threshold at which the equation of state is no longer valid for such a compact object. Thus, these results are likely due to the non-applicability of the EoS in this regime for our theory.

Finally, Fig.\ \ref{fig:massradius_dilaton} shows the mass radius-relation for the dilaton. 
{We likewise adopt the parameter choices of Ref.\ \cite{universe11050158} in order to facilitate a direct comparison with previous white-dwarf calculations. These models are used to illustrate the impact of finite temperature effects in screened scalar-tensor theories and should not be interpreted as a survey of the full observationally allowed parameter space.}
The parameters imposed are ${\tilde\varphi_d = 0}$, ${V_0\, [1] \in 3.3211 \times [10^{-3}, 1]}$ or ${V_0\, [{M_P^2}] \in [10^{-91}, 10^{-88}]}$,\footnote{An unresolved parameter issue was encountered when reproducing the results of \cite{universe11050158} for this theory. The reported mass-radius relations are recovered only for parameter values smaller by a factor of $10^{-3}$ than those obtained from the theoretical conversion. Independent checks did not identify the source of this discrepancy. As the resulting stellar models agree exactly with those of \cite{universe11050158}, we adopt the values given here. Our numerical implementation uses the dimensionless values $V_0[1]$.} and ${\log_{10} a_2 \in [1, 4]}$. {Note that solar system tests suggest $a_2 \gtrsim 10^6$ \cite{Brax_2017} and $V_0 \lesssim 10^{-120} M_P^4$ for compatibility with the current vacuum energy
density \cite{PhysRevD.69.044026}.} Again, the same trend as before holds. At higher temperatures, the white dwarf can hold more mass at the same radius or, equivalently, the radius of the star is larger for the same mass compared to the zero temperature results. We also observe the convergence of the $T=10^8K$ curves at larger radii. In both the symmetron and dilaton case, the mass-radius relation seems to converge towards a mass of approx.\ $0.75 M_\odot$.

\begin{figure*}
    \centering
    \subfloat[[Massive Brans-Dicke theory with $\alpha_0 = 1$ and dimensionless field masses ${m_{\tilde{\varphi}}[1] \in [1\times 10^{4}, 2 \times 10^{5}]}$. \label{fig:Geff_BD}]{
        \includegraphics[width=0.45\textwidth]{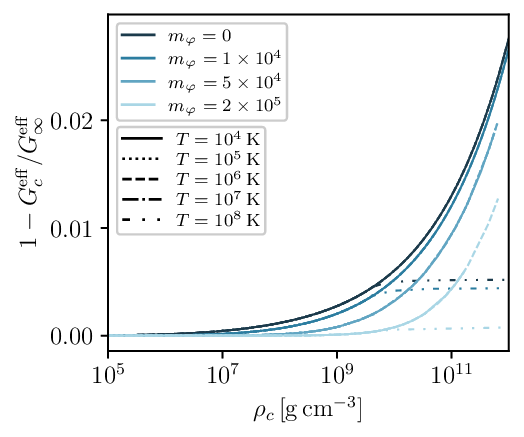}
    }
    
    \medskip

    \subfloat[Symmetron Screening for ${\mu [1] \in \{1.8, 5.4, 9.0\}\times 10^{3}}$, ${\bar{M}_S=10^{-2}}$, and ${\sqrt{\lambda} = \sqrt{2}\mu[1]/\bar M_S^2}$. \label{fig:Geff_symmetron}]{
        \includegraphics[width=0.45\textwidth]{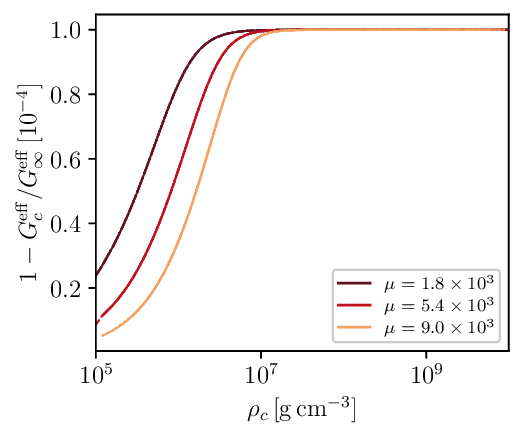}
    }\hfill
    \subfloat[Dilaton screening with $\tilde\varphi_d = 0$, $\log_{10}(a_2) \in \{1, 2, 3, 4\}$ and ${V_0 [1] \in 3.3211 \times [10^{-3}, 1]}$. \label{fig:Geff_dilaton}]{
        \includegraphics[width=0.45\textwidth]{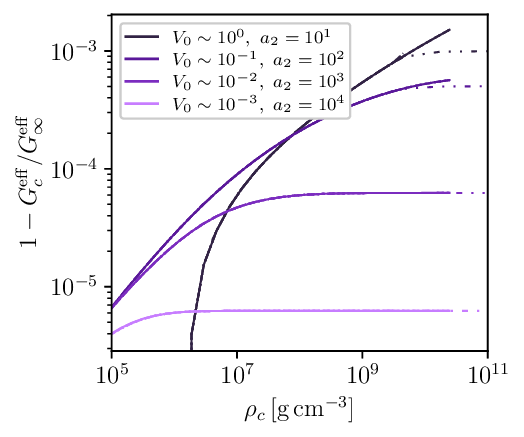}
    }

    \caption{The central effective gravitational constant $G_\mathrm{eff} = A^2(\tilde\varphi)G$ as a function of central density for WDs in Jordan frame for massive Brans-Dicke theory (top), symmetron (bottom left) and dilaton screening (bottom right). Results are presented for temperatures $T \in [10^4, 10^8]K$. All plots follow the temperature linestyle legend presented in Subfig.\ \ref{fig:massradius_GR}.}
    \label{fig:Geff}
\end{figure*}

\subsection{Effective gravitational constant} \label{subsec:results_Geff}

When transforming the action \eqref{eq:action} back to the Jordan frame, we can define an effective gravitational constant that depends on the local value of the scalar field
\begin{equation}
    G^\mathrm{eff}(\tilde\varphi) = A^2(\tilde\varphi)G\;.
\end{equation}
{This quantity controls the strength of gravity in the sense of how strongly the physical energy-momentum of matter curves the Jordan frame (physical) metric via the Einstein equation. It can be distinguished from the effective Newton's constant which is defined in the parametrized post-Newtonian (PPN) framework for slowly moving test particle in weak field  approximation \cite{Perivolaropoulos:2009ak,Hohmann:2013rba,Scharer:2014kya,Jarv:2014hma}. The latter is valid in spatial asymptotics, but does not hold inside a star. In order to obtain an intuitive understanding whether strength of gravity is enhanced or suppressed in a stellar body we compare}
%We denote 
the effective gravitational constant at the center of the star, ${G^\mathrm{eff}_c := G^\mathrm{eff}(\tilde\varphi_c)}$, and at spatial infinity, ${G^\mathrm{eff}_\infty := G^\mathrm{eff}(\tilde\varphi_\infty)}$.

\begin{figure*}
    \centering
    \subfloat[Scalar field \label{fig:scalarfieldprofile_STT}]{
        \includegraphics[width=0.45\textwidth]{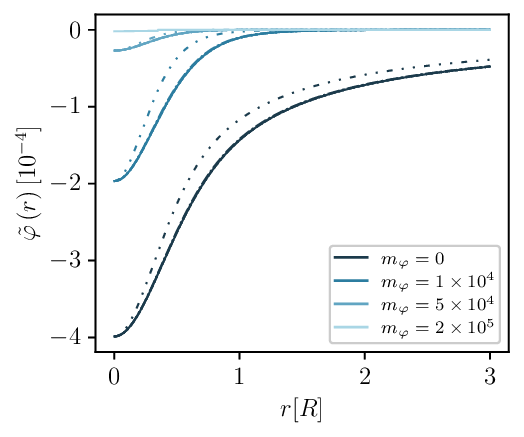}
    }\hfill
    \subfloat[Scalar field derivative \label{fig:scalarfieldderivativeprofile_STT}]{
        \includegraphics[width=0.45\textwidth]{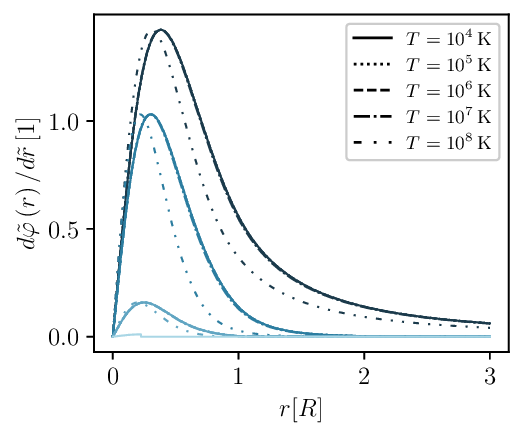}
    }

    \medskip

    \subfloat[Pressure \label{fig:pressureprofile_STT}]{
        \includegraphics[width=0.45\textwidth]{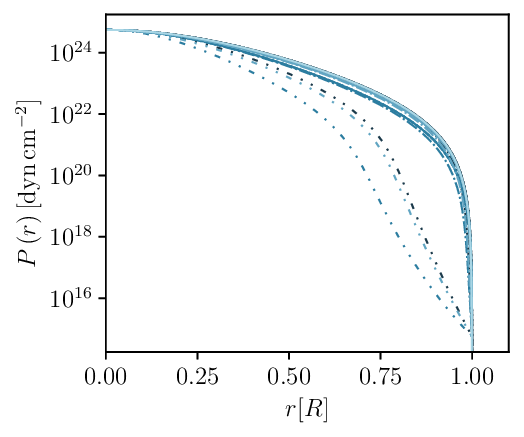}
    }\hfill
    \subfloat[Radial metric function \label{fig:lamdaprofile_STT}]{
        \includegraphics[width=0.45\textwidth]{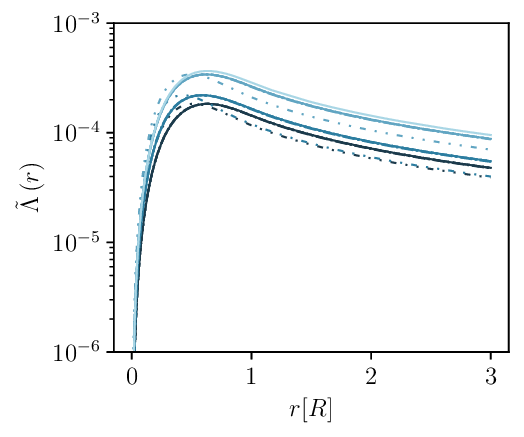}
    }
    
    \caption{Radial stellar profiles in massive Brans-Dicke theory for a star with ${P_c \approx 5.55 \times 10^{24}\, \mathrm{dyn}\, \mathrm{cm}^{-2}}$. All figures follow the color legend given in Subfig.\ \ref{fig:scalarfieldprofile_STT} and temperature linestyle from Subfig.\ \ref{fig:scalarfieldderivativeprofile_STT}. {Pressure $P$ and radius $r$ are Jordan frame quantities, while the scalar field $\tilde\varphi$, its derivative $\d\tilde\varphi/\d\tilde r$ and radial metric $\tilde\Lambda$ are presented in Einstein frame.}}
    \label{fig:radialprofiles_BD}
\end{figure*}

Fig.\ \ref{fig:Geff} shows the deviation of the effective gravitational constant $G^\mathrm{eff}_c$ at the center of the white dwarf from its value $G^\mathrm{eff}_\infty$ at infinity for each theory considered.
We see that for all three theories, the scalar field effectively reduces the strength of gravitation. Two changes in the stellar configuration result from this reduction. First, at the same mass, we find a smaller stellar radius. This might seem counter intuitive at first glance, but can be explained by having a closer look at the hydrostatic equilibrium equation \eqref{eq:hydrostaticequilibrium}. For GR, the scalar field gradient is zero, $\tilde\varphi' = 0$, as no field is present. Further, the first parenthesis and $\tilde\phi'$ are positive, leading to a negative pressure gradient $P'$. This is of course needed to reduce the pressure towards the stellar surface. Now, when adding a scalar field, we get an additional contribution from the term $\alpha\tilde\varphi'$. In all three considered modified gravity theories, we have a positive value of $\alpha(\tilde\varphi)$ and positive scalar field gradient throughout the star as can be seen, e.g., in Fig.\ \ref{fig:scalarfieldderivativeprofile_STT}. Including a field results in a larger absolute value of the pressure gradient. Thus, the pressure drops faster with increasing radius, making the star smaller.
Second, stars with the same mass have a higher central pressure in scalar-tensor theories. This is also physically plausible. As the radius is reduced, a higher central density is needed, and therefore also higher central pressure, to obtain the same mass.

Let us turn toward massive Brans-Dicke theory, for which the results are shown in Fig.\ \ref{fig:Geff_BD}. Here, we have $\tilde\varphi_\infty = 0$ and thus ${G^\mathrm{eff}_\infty = {G}}$, {the bare}
gravitational constant {as it features in general relativity}. First, we observe that this theory shows the largest deviations between the effective gravitational constant at the center of the star and its value at infinity among all the considered modified theories, in the order of a few percent. This effect is also reflected in the mass-radius relation discussed above. Within the theory, the difference increases towards higher central densities. Again, this is consistent with the deviations we observe in the mass-radius relation in Fig.\ \ref{fig:massradius_BD}, as large central densities correspond to larger masses and smaller radii. Finally, we see that the discrepancies are suppressed with larger scalar field mass. We do not observe a significant temperature effect on the effective gravitational constant except for very high temperatures. For ${T=10^8K}$, the difference between $G^\mathrm{eff}_c$ and $G^\mathrm{eff}_\infty$ seems to saturate at high central densities for all scalar field masses. This is probably either related to the validity of the EoS at high temperatures or simply a numerical issue.

Contrary to massive Brans-Dicke theory, the effective gravitational constant of the symmetron does not converge towards {$G$} %$G_N$ 
at infinity. In this theory, we have ${\tilde\varphi_\infty = \mu/\sqrt{\lambdabar} = \bar{M}_S^2}/\sqrt{2}$, such that
\begin{equation*}
     G^\mathrm{eff}_\infty =  \left(1 + \frac{\mu^2}{\lambdabar\bar{M}_S^2}\right)^2{G}%G_N 
     = \left(1 + \frac{\bar{M}_S^2}{2}\right)^2 {G}%G_N
     \;.
\end{equation*}
Now, looking at Fig.\ \ref{fig:Geff_symmetron}, we see that the deviations {in the center from the asymptotic value}
are also much smaller, {$O(10^{-4})$}, and we do not observe any significant temperature effect at all. Nevertheless, an interesting effect is observed at high central densities: the deviation from $G^\mathrm{eff}_\infty$ reaches a plateau. This is an effect of the effective potential \eqref{eq:Symm_potential} of the theory. For high enough central densities, the density term dominates over $\mu$ and the field sits at zero, the potential minimum which no longer depends on the density. At this point $G^\mathrm{eff}_c = {G}%G_N
$ and the theory is equivalent to GR. This can also be seen in the mass-radius plot, Fig.\ \ref{fig:massradius_symmetron}, as all theories converge towards GR at small radii and high masses. Note that the difference is the largest in this region since we calculate it between the central value and the one at infinity. We have
\begin{equation*}
    1 - G^\mathrm{eff}_c/G^\mathrm{eff}_\infty \approx 1 - \left(1 + \frac{\bar{M}_S^2}{2}\right)^{-2} \approx 10^{-4} \;.
\end{equation*}

Finally, the dilaton mechanism is presented in Fig.\ \ref{fig:Geff_dilaton}. In this case, the ratio between the central value of the effective gravitational constant and it's value at infinity is
\begin{equation*}
    G^\mathrm{eff}_c/G^\mathrm{eff}_\infty = \left(\frac{1 + a_2 \tilde\varphi_c^2}{1 + a_2 \tilde\varphi_\infty^2}\right) = \left(\frac{1 + a_2 \tilde\varphi_c^2}{1 + 1/32a_2}\right)
\end{equation*}
using $\tilde\varphi_\infty = \sqrt{2}/{8a_2}$. At lower central densities, the star is less screened so that the central scalar field value goes to its asymptotic value. Hence, $G^\mathrm{eff}_c \to G^\mathrm{eff}_\infty$ and $1-G^\mathrm{eff}_c/G^\mathrm{eff}_\infty \to 0$, which is why we see a drop off towards lower densities for all theory parameters in Fig.\ \ref{fig:Geff_dilaton}. On the other side, at larger central densities, the field is screened more effectively so that the difference between $G^\mathrm{eff}_c$ and $G^\mathrm{eff}_\infty$ rises. Again, we almost see no temperature effects on the effective gravitational constant except for $T = 10^8 K$. For the pairs of theory parameters $\{V_0, a_2\} = \{3.32, 10\}$ and $\{0.33, 10^2\}$, we observe a saturation value for $G^\mathrm{eff}_c$ at high central densities  similar to the massive Brans-Dicke case. Again, we remind the reader that this is likely to be a numerical or EoS validity issue.

\subsection{Interior stellar profiles}

In Fig.\ \ref{fig:radialprofiles_BD} we present an example of how physical quantities change within the star under the assumption of different temperatures for the equation of state. Since we observed a similar behavior for all three theories, we restrict the presented results to massive Brans-Dicke theory and a star with a central pressure of ${P_c \approx 5.55 \times 10^{24} \mathrm{dyn}}$.\footnote{This is not equivalent to the same central density for all models; remember that the relation between pressure and density is temperature dependent (see Fig.\ \ref{fig:EOS})} In these plots, the scalar field mass follows the color legend in Fig.\ \ref{fig:scalarfieldprofile_STT}, while the temperature is represented by the different linestyles shown in Fig.\ \ref{fig:scalarfieldderivativeprofile_STT}. We further normalize the radial coordinate to the stellar radius for each shown model, respectively, meaning that ${r=1}$ corresponds to the stellar surface.

First, we present the radial profile of the scalar field in Fig.\ \ref{fig:scalarfieldprofile_STT}. We note that, as expected, the magnitude of the field is strongest at the center of the star and subsequently decreases towards spatial infinity. In addition, the higher the mass of the field, the stronger and radially faster the suppression. This suppression is further intensified at very high temperature, i.e.\ $T=10^8 K$. Otherwise, we do not observe any temperature effects when considering a normalized radius.

We continue to present the scalar field derivative in Fig.\ \ref{fig:scalarfieldderivativeprofile_STT}. In consistence with the scalar field profile, we observe that the field's derivative grows to a higher maximum and converges more slowly towards zero for lower field masses. For ${T=10^8 K}$, the initial increase and later conversion towards zero occurs already at smaller radii compared to lower temperatures. Again, we do not see any other temperature effects on the radial profile of the scalar field derivative.

Fig.\ \ref{fig:pressureprofile_STT} depicts the radial pressure profile of the star for all the considered model parameters. Here, we already observe a slight deviation from the pressure profile at zero temperature for ${T=10^7K}$. The pressure declines faster compared to $T < 10^7K$. For ${T=10^8K}$, we do not only see an even quicker drop of the pressure, but also a change in shape of the radial profile.

Lastly, we present the radial metric function in Fig.\ \ref{fig:lamdaprofile_STT}. Again, we only observe a deviation from the zero temperature case for ${T=10^8K}$. At this temperature, the radial metric function $\Lambda(r)$ rises and reaches its maximum more rapidly, but also its recession towards zero going to spatial infinity is faster.

Overall, we can summarize that stellar properties are mostly left unchanged with increasing temperature. The only significant deviations arise towards ${T=10^8K}$. As mentioned before, we have to be careful with results for such high temperatures as the gas becomes less degenerate in this regime, and the EoS starts loosing its validity.

\section{Conclusions} \label{sec:conclusions}

In this article, we analyzed how a finite temperature Chandrasekhar equation of state for white dwarfs affects their equilibrium and internal structure in modified gravity theories. We considered a massive Brans-Dicke theory as well as the symmetron and dilaton screening mechanisms.

First, we derived the Chandrasekhar equation of state from kinetic theory considerations. We presented the finite- and zero temperature equations that relate the pressure to the number density of relativistic, degenerate matter. We continue to introduce the field equations for a general scalar-tensor theory. These theories are fully characterized by choosing the conformal factor $A(\tilde\varphi)$ and the self-interacting potential $V(\tilde\varphi)$ of the scalar field. Assuming a spherically symmetric and static setup as well as with the addition of the hydrostatic equilibrium equation and Chandrasekhar equation of state, we arrive at a closed system of differential equations. Physical considerations then set the boundary conditions for each theory, allowing for numerical solutions of the system.

Hereafter, we introduce the theories considered. Massive Brans-Dicke theory is characterized by a constant non-minimal coupling parameter $\alpha_0$ and an effective mass $m_{\Tilde{\varphi}}$ defining a quadratic potential. Screening mechanisms on the other hand, feature an environment dependent effective potential that dictates the behavior of the field. For symmetron screening, we assume a coupling of the form \eqref{eq:Symm_conffactor} and have a Mexican hat like potential \eqref{eq:Symm_potential}. While the field is driven to zero in high density regions, symmetry breaking causes it to converge towards the potential minimum at low densities. In the dilaton case, it is the conformal factor \eqref{eq:Dilaton_conffactor} that directs the field to its minimum $\tilde\varphi_d$. The potential follows an exponential runaway form \eqref{eq:Dilaton_potential}.

After a brief description of our numerical setup, we presented our results starting with the mass-radius relations shown in Fig.\ \ref{fig:massradius}.  We observed a qualitatively consistent pattern for all considered theories. Deviations from the zero temperature result remain very small for temperatures up to $\sim 10^6K$. For $T>10^6K$ deviations increase significantly and can no longer be neglected. We note that for higher temperatures, we obtain a larger radius for the same mass while the central density remains mostly unchanged. This can be easily explained by the increasing thermal pressure that counteracts the gravitational pull. The stellar matter is pushed further out. This result is especially relevant for computations of white dwarf cooling. Cooling time is commonly defined as the time it takes for a white dwarf to cool down from $10^8$ to $10^6K$. Thus, assuming that the stellar radius remains constant during this process is evidently unrealistic. The finite temperature deviations also result in partial degeneracies between different sets of theory parameters for each modified gravity theory which has to be taken into consideration when constraining theory parameters from observations. Lastly, we note that the mass-radius curves for ${T=10^8K}$ seem to converge at large radii and for smaller masses. This hints at the idea that temperature effects could dominate over modifications of the underlying gravity theory in this regime. Again, we emphasize that the finite temperature Chandrasekhar equation of state may well be reaching its non-validity region at this temperature. {We also stress that these models are intended to lay out the effect of finite temperature on equilibrium configurations. A systematic exploration of observationally viable parameter regions for both screening mechanisms is beyond the scope of the present work and is left for future investigation.}

We continued to compute and discuss the modified gravity and temperature effects on $G^\mathrm{eff}$, the effective gravitational constant. We presented our results in Fig.\ref{fig:Geff}. We observed that for all theories considered, the effective gravitation is weaker at the center of the star compared to its value at infinity. Further, we saw the largest deviations for massive Brans-Dicke theory, {$O(10^{-2})$}, which is also reflected in the larger deviations in the mass-radius curves.
For massive Brans-Dicke theory, the only visible temperature effects occurs at large central densities and only for ${T=10^8K}$. The difference in gravitational strength seems to saturate. In the symmetron scenario, we clearly see the effect of the density dependent effective potential. At large enough central densities, the field sits at zero, independently of the specific value of $\rho_c$, which leads to a plateau in the plot of relative gravitational strength. We do not observe any temperature effects for the symmetron. In the dilaton case, we again observe a saturation of the effective grativational constant at high densities for ${T=10^8K}$, similiar to the massive Brans-Dicke case.

Finally, we continued to analyze the radial profile of various physical parameters for a white dwarf with a central pressure of ${P_c \approx 5.55 \times 10^{24} \mathrm{dyn}}$. As we observed that the quantitative behavior remains unchanged between the different theories considered here, we only present results for massive Brans-Dicke theory in Fig.\ \ref{fig:radialprofiles_BD}. The parameters considered turn out to be mostly temperature independent. Significant deviations emerge only for very high temperatures ${T \sim 10^8K}$.

Overall, we can say that the main effect of temperature considerations is the increase in radius with rising temperature. Otherwise, all physical quantities retain their qualitative behavior mostly except for extremely high temperatures. 
As mentioned various times, the results at these extreme temperatures have to be taken with care as we approach the validity limit of the equation of state. A more realistic model would require consideration of the presence of non-degenerate matter in the outer layers of these stars as well as for hotter white dwarfs in general. 
Moreover, at temperatures as high as $T\sim10^{8},\mathrm{K}$, a complete evolutionary description of a white dwarf would require additional microphysical processes beyond the finite-temperature equation of state considered here. In particular, neutrino emission or residual or shell nuclear burning may become relevant.

{
A quantitative confrontation with observations is left for future work, as it would require a dedicated treatment of observational uncertainties and the envelope physics connecting the white dwarf interior to observable quantities. Here, our aim is instead to isolate and characterize the interplay between finite-temperature effects and modifications induced by the scalar field, rather than to constrain the underlying gravity models with observational data.}

\section*{Acknowledgements}
The authors acknowledge the support provided by the COST Action FuSe (CA24101). LJ was supported by the Estonian Research Council team grant ``Space - Time - Matter'' (PRG2608) as well as Estonian Ministry of Education and Research Centre of Excellence ``Foundations of the Universe'' (TK202U4).

{\section*{Data Availability Statement}}

{The data underlying the figures and numerical results presented in this work are available from the corresponding author upon reasonable request.}

\appendix

\section{Theory notation} \label{append:notation}

In the usual notation used for screening mechanisms, the action in Einstein frame is written as

\begin{align}\label{eq:action_screening}
    S =& \frac{c^2}{\hbar}\int \d^4x \sqrt{-\tilde{g}} \left[\frac{M_P^2}{2}\tilde{R} - \frac{1}{2}\tilde{g}^{\mu\nu}\tilde\partial_{\mu}\tilde{\phi}\tilde\partial_{\nu}\tilde{\phi}  - \frac{c^2}{\hbar^2}\mathcal{V}(\tilde{\phi})\right] \nonumber \\
    &+ S_\mathrm{m}\left(\mathcal{A}^2(\tilde{\phi})\tilde{g}_{\mu\nu}, \chi\right)\;,
\end{align}
with restored units, the Planck mass $M_P = \sqrt{\hbar c / 8\pi G}$ and the action having dimensions of energy multiplied by time, i.e. $[S] = [\hbar] = ML^2/T$.
{Comparing the Einstein-frame action used in the screening-mechanism literature \eqref{eq:action_screening} with our Einstein-frame action \eqref{eq:action} one finds that the gravitational sectors are already written in the same form. The remaining differences arise from the normalization of the scalar field and self-interacting potential. Equating the kinetic terms of the scalar field therefore determines the relation between the scalar fields, while equating the potential terms fixes the conversion between the two potential conventions. This yields the relations}

\begin{subequations}
    \label{eq:notation_relations}
    \begin{align}
        \tilde\phi &= \sqrt{\frac{\hbar c}{4\pi G}}\,\tilde\varphi = \sqrt{2}M_P \,\tilde\varphi \;, \label{eq:notation_relations_field} \\
        \mathcal{A}(\tilde\phi) &= A\left(\sqrt{\hbar c/4\pi G}\,\tilde\varphi\right) = A\left(\sqrt{2}M_P \,\tilde\varphi\right) \label{eq:notation_relations_conffactor} \\
        \mathcal{V}(\tilde\phi) &= \frac{\hbar c}{16\pi G} V\left(\sqrt{\hbar c/4\pi G}\,\tilde\varphi\right) \nonumber \\
        &= \frac{M_P^2}{2} V\left(\sqrt{2}M_P \,\tilde\varphi\right)\;. \label{eq:notation_relations_potential}
    \end{align}
\end{subequations}
{The interpretation of these relations is straightforward. In the screening convention, the field $\tilde\phi$ carries mass dimension one, whereas our field $\tilde\phi$ is dimensionless. The conformal factor is dimensionless in both formulations and therefore only its argument changes under the field redefinition. Likewise, the two potential functions describe the same physical self-interaction but are expressed in different conventions. Consequently, the potential acquires the overall factor $M_P^2/2$ appearing in \eqref{eq:notation_relations_potential}. All symmetron and dilaton expressions used in this work follow from these three substitutions.}

\section{Shooting Method} \label{append:shootingmethod}

\begin{figure}
    \centering
    \includegraphics[width=\linewidth]{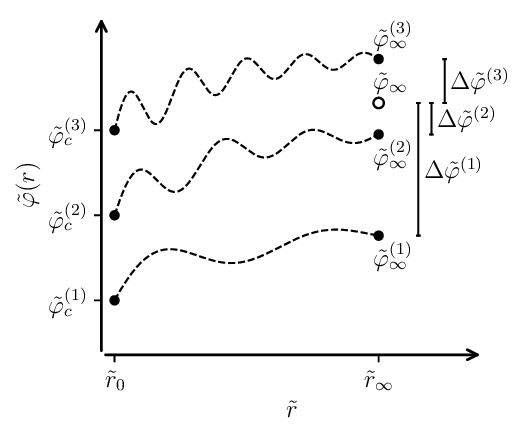}
    \caption{Visualization of the shooting mehtod used to find the initial condition for the scalar field that leads to the correct asymptotics.}
    \label{fig:shootingmethod}
\end{figure}

A system of ordinary differential equations (ODEs) can only be numerically  solved as an initial value problem. In this case, initial conditions are given at the same point, e.g.\ $r=0$, and evolved by integrating the ODE system using, e.g.\, a Runge-Kutta method.
In our case, the boundary conditions are given at the center of the star, $r=r_0$, and towards infinity, $r=r_\infty$. Numerically speaking, $r_0$ and $r_\infty$ are small/large enough numbers respectively \footnote{"Small/large enough" meaning that numerical results will not significantly change if these values are decreased/increased.}. A shooting method reduces this problem to an initial value problem. For variables with a boundary condition at infinity, the method finds the initial value at the center that correctly leads to the given boundary value. These variables are called \textit{shooting parameters}. For the system \eqref{eq:FE_sphsymm}, the scalar field is the only relevant shooting parameter as the temporal metric function $\tilde\phi(\tilde r)$ only enters the equations through its derivative and can thus be eliminated from the equations.

Here we outline the procedure for the scalar field $\tilde\varphi$. A visualization is shown in Fig.\,\ref{fig:shootingmethod}. The method follows the following procedure:
\begin{enumerate}
    \item First, we set the target value of the scalar field at numerical infinity $\tilde\varphi_\infty$ and the score function
    \begin{equation*}
        \Delta\tilde\varphi := \tilde\varphi_\infty - \tilde\varphi(\tilde{r}_\infty)\,.
    \end{equation*}
    The score function is a directional measure for how far off we are from our desired target value and $\Delta\tilde\varphi = 0$ means that we exactly hit our target.

    \item The first iteration takes a random guess for the initial value of the scalar field $\tilde\varphi_c^{(1)}$. We evolve our variables by integrating the differential equations with this guess and the remaining initial conditions. In this way, we find the value of the scalar field at infinity $\tilde\varphi_\infty^{(1)}$ and can compute the score function $\Delta\tilde\varphi^{(1)}$.
    \item In the next step, the code slightly varies the first initial guess to define a second initial guess $\tilde\varphi_c^{(2)}$. Again, the variables are evolved using the ODEs. We find $\tilde\varphi_\infty^{(2)}$ at infinity and can compute $\Delta\tilde\varphi^{(2)}$.
    Let us assume that $\tilde\varphi_c^{(2)}>\tilde\varphi_c^{(1)}$ and $\sgn(\Delta\tilde\varphi^{(2)}) = \sgn(\Delta\tilde\varphi^{(1)})$. If $|\Delta\tilde\varphi^{(2)}| > |\Delta\tilde\varphi^{(1)}|$, we are moving in the wrong direction. Our third guess should be smaller than $\tilde\varphi_c^{(1)}$. If $|\Delta\tilde\varphi^{(2)}| < |\Delta\tilde\varphi^{(1)}|$ and $\Delta\tilde\varphi^{(2)} \neq 0$, we are moving in the right direction, but still haven't hit our target value. Let us assume the latter case and continue to a third guess.
    \item Our assumption in the last step means that our third guess should be larger than the previous ones, i.e. $\tilde\varphi_c^{(3)}>\tilde\varphi_c^{(2)}$. Once more, we solve the system with this guess and compute the score function. Let us now assume that $\sgn(\Delta\tilde\varphi^{(3)}) \neq \sgn(\Delta\tilde\varphi^{(2)})$. This means that we are beyond our target value $\tilde\varphi_\infty$, we \textit{overshooted}. Hence, the desired initial value must lay between the second and third guess, i.e. $\tilde\varphi_c^{(3)}>\tilde\varphi_c>\tilde\varphi_c^{(2)}$.
    \item By carrying on with this procedure, we can continue to narrow down the correct initial value for the scalar field $\tilde\varphi_c$ that hits the target value at infinity $\tilde\varphi_\infty$.
    Note that numerically speaking, we will never arrive at $\Delta\tilde\varphi = 0$. Instead, we have to set the desired accuracy, e.g.\ $|\Delta\tilde\varphi| = 10^{-10}$. Once the score function goes below this value, we accept the result as ``correct."
\end{enumerate}

\bibliographystyle{utphys}
\bibliography{biblio}

\end{document}